\documentstyle[12pt,epsfig]{article}
\newcommand{\be}{\begin{equation}}
\newcommand{\ee}{\end{equation}}
\newcommand{\ba}{\begin{eqnarray}}
\newcommand{\ea}{\end{eqnarray}}

\newcommand{\fr}[2]{\frac{#1}{#2}}

\newcommand{\lb}{\left(}
\newcommand{\rb}{\right)}
\def\vec#1{{\mbox{\boldmath$#1$}}}
\newcommand{\r}{\mbox{$\vec{r}$}}

\begin{document}

\thispagestyle{empty}
\begin{flushright}
WUE-ITP-98-010\\
April 1998
\end{flushright}
\vspace{0.5cm}
  \begin{center}
{\Large \bf Top quark production near threshold:}\\[.3cm]
{\Large \bf on resummation of
${\cal O}\Bigg( (\beta_0\alpha_s)^n\Bigg)$
QCD corrections}  \\
\vspace{1.7cm}
{\sc \bf O.~Yakovlev$^{a}$ and A.~Yelkhovsky${^b}$}\\[1cm]
\begin{center} \em
 $^a$ Institute f\"ur Theoretische Physik, Universit\"at W\"urzburg,\\
D-97074 W\"urzburg, Germany \\
\vspace{4mm}
$^b$Novosibirsk University and Budker Institute of Nuclear Physics,\\
Novosibirsk, 630090, Russia
\end{center}\end{center}
\vspace{1cm}
\begin{abstract}
We discuss the resummation of the potentially large
${\cal O}\Big( (\beta_0\alpha_s)^n\Big)$
QCD corrections to the total cross section of the process
$e^+e^-\to t \bar t$ near the threshold.
In this approximation, the cross section factorization
into the short-- and  long--distance parts is valid.
The short--distance correction is reduced to the production vertex
renormalization.
It amounts to $-9.5\%$  and is well under
control.
The long distance corrections are accounted for as the effect of
the coupling's running in the QCD potential.
We argue that the accuracy of present predictions
for the cross section in the scheme with the running QCD coupling
is about 10\%.
\end{abstract}
\centerline{\em PACS numbers: 12.38.Bx, 12.38.Cy, 13.85.Lg, 14.65.Ha}
%\noindent Keywords: top quark, inclusive cross section, perturbative
%calculations.

\vfill

\noindent $^a${\small e-mail: iakovlev@physik.uni-wuerzburg.de} \\
$^b${\small e-mail: yelkhovsky@inp.nsk.su}

\newpage

{\bf 1. }
The theoretical studies of the process $e^+e^-\to t\bar t$ near the
threshold
are of interest for many applications. First of all,
the top quark will be studied in the near threshold kinematics at
the Next Linear Collider, and precise measurements of its mass and width
are among the most important experimental issues \cite{topphys}. From this
standpoint, reliable estimations of the QCD perturbative effects
are mandatory.

The general approach for the problem of the top quark
production near the threshold was suggested in \cite{FK},
where it was demonstrated that the top quark width plays
the role of an infrared cutoff which suppresses
all nonperturbative long--distance effects. The size of the nonperturbative
corrections was estimated in \cite{SP,FY1} to amount to
less than one
per cent. Therefore, one could hope to
predict the total cross section by perturbative calculations with such an
accuracy.
 One--loop QCD corrections
were discussed in \cite{SP,OneL,Green} and
Higgs--induced electroweak corrections,
which have the same order as the QCD ones,
were discussed in \cite{Higgs}.
Recently, the ${\cal O}(\alpha_s^2)$ QCD perturbative corrections to the
$t\bar t$ threshold cross section were calculated in \cite{HT,MY}
and were found to be comparable in size with the first order correction.
This result can cast some doubts upon
the applicability of QCD perturbation theory to the problem under
discussion.

In this note we make an attempt to answer the question:
``How can
higher order perturbative corrections modify the NNLO  result?".
Surely, an exact answer to this question deserves an exact knowledge of
those higher order corrections
which is far beyond our reach at the moment.
Nevertheless, there exists a method to estimate a contribution
of higher orders, which is known to work well at different problems 
(see for example \cite{BG,N95,Braun,BB,BBB,Dok,Ural1} and references 
therein). 
This method exploits the large value of
$\beta_0=11-\frac{2}{3}N_F$, the lowest order coefficient from the QCD
beta--function, and consists in an approximation of the
complete result for
the $(n+1)$th order, by its counterpart of the $n$th order in $\beta_0$.
The latter
is found by the naive non-abelianization (NNA) prescription: in the
(gauge--invariant) result for a set of Feynman diagrams containing $n$
light quark vacuum polarization bubbles, one replaces $N_F^n$ by
$(-3\beta_0/2)^n$. In other words, taking into account only the running
of the QCD coupling, one assumes it to be a reasonable approximation for
all radiative corrections.

A nice feature of this approach with respect to the threshold cross section
analysis is a complete factorization of hard and soft corrections, both  of
which reside on its own scale.

In what follows, we first discuss the intermediate kinematic region
$\alpha_s\ll v \ll 1$.
We resum  ${\cal{O}}\Big( (\beta_0\alpha_s)^n\Big)$
 terms to the first two leading coefficients
in the expansion in $v$.
Then, we consider the cross section near the threshold,
where short--distance effects are factorized to the normalization
of the cross section and the long--distance effects are absorbed to
the static potential with the running coupling.
We perform a numerical evaluation with such the potentials to
test the influence of the higher order QCD corrections on
the cross section.

{\bf 2. }
Let us first consider the kinematic region $\alpha_s \ll v \ll 1$, where
$v $ is the
relative velocity of quark and antiquark. In this region, we have an
extra small parameter, $\alpha_s/v$, so that the interaction between the
slowly moving quark and antiquark can be treated perturbatively. In the
lowest order, this interaction is caused by the one--gluon exchange between
two particles. Corresponding one--loop correction to the pair production
amplitude, $A_0(v)$, can be easily obtained using the QED result
by Schwinger \cite{Schw}:
\be\label{A0}
A_0(v) = \frac{C_F\alpha_s}{\pi} \left\{ \fr{ \pi^2 }{ 2v } - { 2 }
\right\}.
\ee
Here the former term is determined by the low--energy scale, so that all
powers of $C_F\alpha_s/v$ should be resummed in the vicinity of the
threshold. The latter term
has its origin at the relativistic scale.

As discussed above, we would like to consider the effects of
vacuum polarization only. The most appropriate way to do that is to
substitute $1/(k^2 - \lambda^2)$ for $1/k^2$ in the gluon propagator,
and to calculate the amplitude $A_{\lambda}(v)$ as a function of
the ``gluon mass" $\lambda$.
Then the first--order in $\alpha_s$ correction to the production
amplitude, which is valid to all orders in $\beta_0\alpha_s$, equals
\cite{Braun,BB,BBB,Dok,N95}
\be\label{A}
A(v) = - \fr{ 1 }{ \pi } \int_{-\infty}^{\infty}
                         \fr{ d\lambda^2 }{ \lambda^2 }
                         \left[ A_{\lambda}(v) - A_0(v) \right]
                         \mbox{Im} \fr{ 1 }{ 1 + \Pi(\lambda^2) },
\ee
where
\be
\Pi(k^2) = - \fr{\beta_0\alpha_s}{4\pi}
            \ln \lb - \fr{ k^2 }{ \mu^2 } e^C \rb
\ee
is the one--loop polarization operator, $\mu$ is the normalization
point for the coupling constant, $\alpha_s \equiv \alpha_s(\mu)$,
while $C$ is a scheme--dependent
constant. In what follows, we use $C_{\overline{\rm MS}}= -5/3$.
The result for $A_{\lambda}(v)$ reads \cite{AY,r1}:
\ba
A_{\lambda}(v) &=&  \fr{ C_F\alpha_s }{ v }
                    \mbox{arctg} \fr{ v }{ z } \label{soft} \\
                    &&  - \fr{2C_F\alpha_s}{3\pi}\left\{\fr{3\pi}{2z}+
                    \frac{z^6-2z^4-2z^2-6}{z\sqrt{z^2-4}}\mbox{arch}
                    \fr{z}{2}-z^4\ln z + z^2 + \fr{3}{2} \right\}
                    \label{hard}\\
               &=&  A^s_{\lambda}(v) + A^h_{\lambda}.
\ea
Here $z \equiv \lambda/m$, $m$ is the mass of the $t$--quark.
Inserting the difference $A_{\lambda}(v) - A_0(v)$ into (\ref{A}), we
obtain
the amplitude in the intermediate kinematic region as a sum of the soft and
hard terms,
\be\label{AA}
A(v) =  A^s(v) + A^h .
\ee

{\bf 3.
}
Now we would like to find the hard correction, $A^h$, due to
(\ref{hard}). Inserting (\ref{hard}) into the integral (\ref{A}), one can
obtain its expansion in power series over $a =
\beta_0\alpha_s(\mu)/(4\pi)$:
\be
A^h = - \frac{C_F\alpha_s(m)}{\pi} \sum_{n=0} r_{n} a^n,
\ee
where the lower order coefficients at $\mu =m$ are:
\be
r_{0} = 2, \qquad r_{1} = \fr{ 11 }{ 6 },
\qquad r_{2} = \fr{ 4\pi^2 }{ 3 } + \fr{ 163 }{ 18 },
\qquad r_{3} \approx 57.05, \qquad r_{4} \approx 1131.
\ee
The zeroth order coefficient coincides with that from Schwinger's result
(\ref{A0}), the first order one is in accord with the result obtained in
\cite{r1}, while $r_{2}$, $r_{3}$ and  $r_{4}$ are new.
Due to the fact that the vector current's anomalous dimension is zero, one
can
easily restore the $\mu$--dependence of the results.
In Fig.1, we compare the results of the lower orders inclusion with the
exact (in NNA approximation) result. The actual values of $a$ for $c$--,
$b$--
and $t$--quarks are indicated in order to demonstrate how the perturbation
theory works for the current renormalization at the corresponding
thresholds.

\begin{figure}
\centerline{
\epsfig{bbllx=100pt,bblly=209pt,bburx=507pt,%
bbury=490pt,file=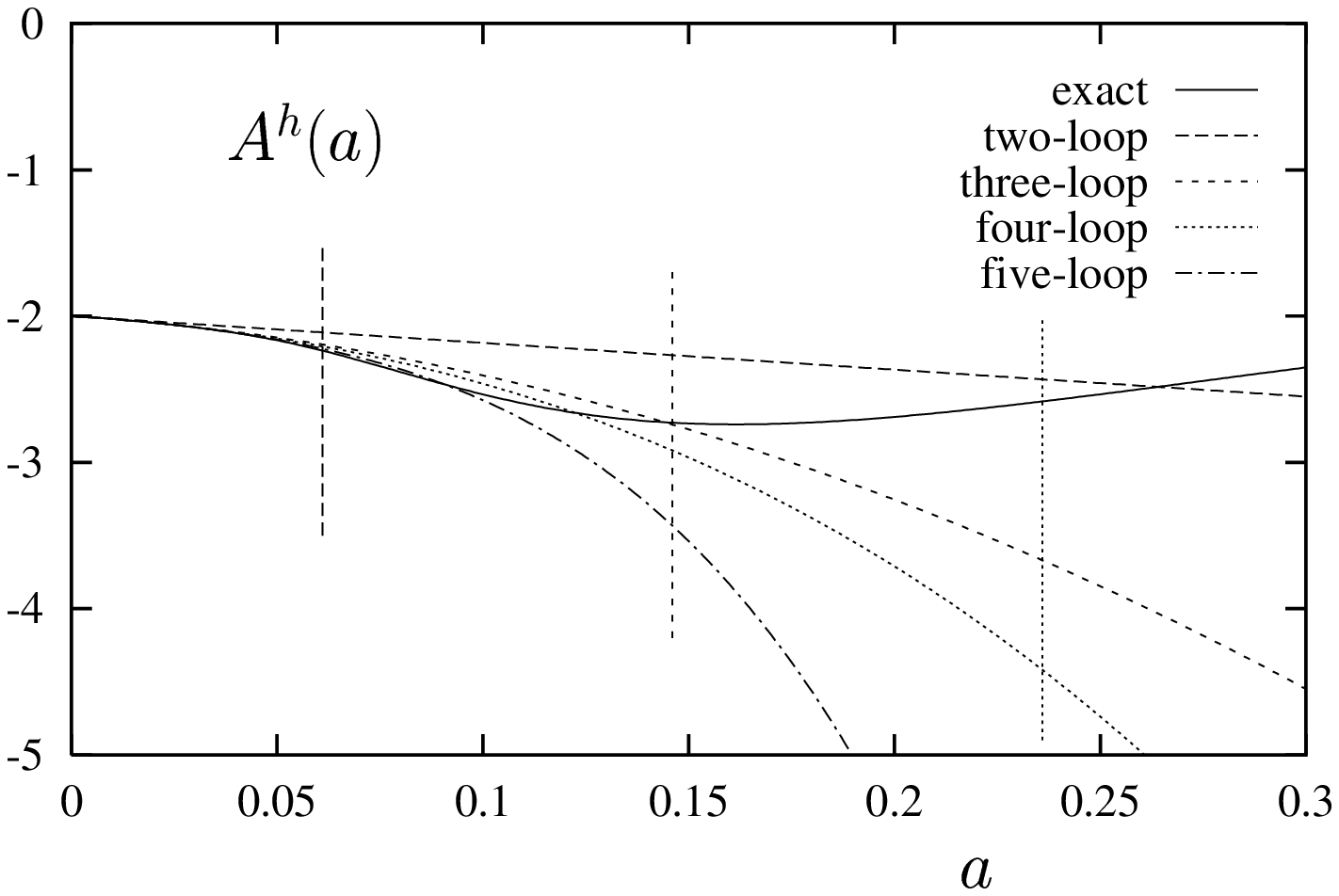,scale=0.9,%
clip=}
}
\caption{\it
 $A^h_{``exact"}(a)$ (solid line);
$A^h$ (dashed and dotted lines) expanded in $a^n$,
 as a function of $a$.
 Vertical lines mark the actual values of $a$ at the threshold
 for top ($m=$175 GeV), bottom ($m=$4.8 GeV), and charm ($m=$1.4 GeV)
quarks.}
\end{figure}

 Taking $\alpha_s(m_t) = 0.1$, so that $a = 0.061$,
we can calculate an
``exact" value of the hard correction to the amplitude of $t\bar t$
production:
\begin{eqnarray}\label{Ah}
A^h_{\rm ``exact"}(a)|_{a\to 0.061} &= & -2.236  \frac{\alpha_sC_F}{\pi}
\\
&\approx&(-2-1.33a-41.7a^2)\frac{\alpha_sC_F}{\pi}.
\end{eqnarray}
The second equation is an extrapolation formula which
works in the region $a=(0,0.1)$
and can be used to obtain
$A^h_{\rm ``exact"}(a)$ at different  $\alpha_s(m_t)$.
Adding up this result to unity, we obtain the renormalization of the vertex
$\gamma^*t\bar t$ at the threshold, by the hard QCD correction (in NNA
approximation).
The ${\cal{O}}(\alpha_s^2\beta_0)$ correction amounts to about $6\%$ of the
one--loop ${\cal{O}}(\alpha_s)$ result $A_0^h$, the ${\cal{O}}(\alpha_s^3
\beta^2_0)$ correction gives about $4\% $
and the rest in the sum gives $ 2\%$.
The sum of higher order corrections constitutes $12\%$ of $A^h_0$.
The net QCD correction to the renormalization of the $\gamma^* t\bar t $
vertex amounts to $-9.5\%$, in the NNA approximation.

We would like also to compare this number with the two-loop QCD
correction  to the vector current normalization constant, $A^h_{\rm
two-loop}(\mu^2_{\rm fact})$, derived in
\cite{CzM,MY,HT}, which depends on a
factorization scale. We take eq.(22) from
\cite{HT} (or eq.(39) from \cite{MY}), to obtain
$A^h_{\rm two-loop}(m^2)=-2.34\frac{\alpha_sC_F}{\pi}$ and
$A^h_{\rm two-loop}(m^2/2)=-2.14\frac{\alpha_sC_F}{\pi}$.
One can see that our result (\ref{Ah}) is rather close to these numbers.

We conclude that the discussed corrections to the short--distance part
of the production amplitude are well under control.

As for the long--distance part in the intermediate region, we obtain
\be
A^s(v)=  \frac{C_F\alpha_s(\mu)\pi}{v}\sum_{n=0} t_{n} a^n,
\ee
where the lower order coefficients are:
\begin{eqnarray}
t_0&=&\frac{1}{2},\quad
t_1=-\frac{ D }{ 2 },\quad
t_2=  \frac{ D^2 }{ 2 } + \frac{\pi^2}{3}, \\ \nonumber
t_3&=& - \frac{ D^3 }{ 2 } - \pi^2 D, \quad
t_4= \frac{ D^4 }{ 2 } + 2D^2\pi^2+\frac{8\pi^4}{5},
\end{eqnarray}
and $D=C-2\ln (\frac{\mu}{mv})$.
The zeroth order coefficient coincides with that
 from Schwinger's result
(\ref{A0}), the first order one coincides with the result obtained in
\cite{r1}, while $t_{2}$, $t_{3}$ and  $t_{4}$ are new.

It is worthy to stress that the intermediate region has only academic
interest, since at $\alpha_s=0.1$ there is no room for the strong
inequality
$\alpha_s\ll v\ll 1$.  Either ${\cal{O}}(v^n)$ relativistic corrections
become equally important, or the ratio $\frac{\alpha_s}{v} $ proves to be
of
order unity.  On the other hand, one can use the results for the short
distance correction $A^h$ directly at the threshold region, after
subtraction
of the resummed Coulomb singularities.  In what follows we discuss such a
resummation.

{\bf 4.
}
For the sake of completeness,
we would like to discuss how
the running of the QCD coupling modifies the cross section near the
threshold.
This issue was a subject of many papers. It was discussed in
\cite{SP,Green},
with the NLO accuracy and in \cite{Kuhn}, with the NNLO accuracy.
Our main task here is to analyze how the perturbative expansion works in
the scheme
with the running QCD coupling and how accurate are the
predictions, made in this scheme.
We perform a numerical analysis in the position space and compare its 
results
with those obtained in the fixed normalization point scheme. Our analysis
is complementary to that made in \cite{Kuhn}, where a comparison of the
potentials
in the momentum and position spaces was discussed.

    Near the threshold, the ratio $\frac{\alpha_s}{v}$
is not small and one has to resum all
${\cal{O}}\Big( (\frac{\alpha_s}{v})^n\Big) $
terms \cite{SS}.
For the $t\bar t$ production, we should also include the
finite width of the $t$--quark, $\Gamma_t$.
The cross section of the process $e^+e^-\to t\bar t$ near the threshold
normalized to the cross section for $e^+e^-\to\mu^+\mu^-$, is\footnote{In
what follows, we disregard relativistic corrections.}
\cite{FK}:
\be
R =  N_c e_t^2 \frac {24\pi}{s} \mbox{Im}\; G(E+i\Gamma_t;0,0),
\ee
where $G(E+i\Gamma_t;\vec{r},\vec{r}')$ is the Green function
for the Schr\"odinger equation:
\begin{eqnarray}
(H-E-i\Gamma_t)G(E+i\Gamma_t;\vec{r},\vec{r'})=\delta( \r-\r'),\quad
H=\frac{\vec{p}^2}{m}+V(r).
\end{eqnarray}
Thus, the problem of the soft corrections resummation to
all orders in $\frac{\alpha_s}{v}$,
reduces to a calculation of $G(E+i\Gamma_t;0,0)$ for
the Schr\"odinger equation with a
heavy quark--antiquark
potential $V(r)$.
In the NNA approximation, the short--distance correction $A^h$ derived at
the
previous section, enters into $R$ through the factor $1+A^h$.

It is well known, that at LO \cite{FK} and NLO \cite{SP}
no choice exists of the normalization point
for the coupling constant, which would result in an appropriately similar
behavior of the threshold cross sections calculated
in the fixed point perturbative potential and the ``running" one,
respectively. This means, that an account of running is actually important,
since it allows one to resum large logarithmic corrections
which arise in the fixed point perturbation theory.
However, an account of the NNLO corrections makes two cross sections much
closer to each other, if the high normalization point is used,  $\mu \sim
m_t$.
The difference becomes an effect of ${\cal{O}}(\alpha_s(\mu)^3)$ and
appears
to be not very large, at the level of 5-10\%.

Let us now discuss a scheme with the running QCD coupling.
Usually a starting point for the analysis is a perturbative potential
in the scheme with a fixed normalization point, in the momentum space.
The ``running" potential is obtained  by resumming all logarithmic
terms
to the running coupling:
\begin{eqnarray}\label{V}
V(r)=&-&C_F\frac{\alpha_s(1/r')}{r}\Bigg\{ 1
+\frac{\alpha_s(1/r')}{4\pi}a_1+\Bigg(\frac{\alpha_s(1/r')}{4\pi}\Bigg)^2
\Big[ \beta^2_0\frac{\pi^2}{3}+a_2 \Big]\\ \nonumber
&+&\Bigg(\frac{\alpha_s(1/r')}{4\pi}\Bigg)^3
\Big[ 16 \beta_0^3\zeta(3)
+\beta_0\lb 3\beta_0a_1
+\frac{5}{2}\beta_1\rb
\frac{\pi^2}{3}+a_3
\Big]
\Bigg\}.
\end{eqnarray}
Here $r'=re^{\gamma_E}$, $\gamma_E$ is the Euler constant.
The coefficients $a_1,a_2$ are known \cite{FB,Peter},
while $a_3$ is still unknown.
The coupling $\alpha_s(1/r')$ suffers from the
Landau pole, appearing as a consequence of extrapolation into
the strong coupling domain of the perturbation theory result .
The usual way to handle this singularity is an introduction of some
model potential, mimicking non--perturbative effects, at the large 
distances
$r>r_0\sim 1/\Lambda_{\rm QCD}$ \cite{SP,Green}.
However, the contribution of the ``non-perturbative"
region to the resulting cross section proves to be extremely small, less
than $0.1-0.3 \%$, due to the large width of the top quark. Therefore this
cross section
can be considered as model--independent.

To find the cross section for various
 potentials, we solve
the Schr\"odinger equation numerically.
Our result for the running NNLO potential agrees with the corresponding
result of \cite{Kuhn}, obtained in coordinate space,
when the same input parameters are chosen.
\begin{figure}
\centerline{
\epsfig{file=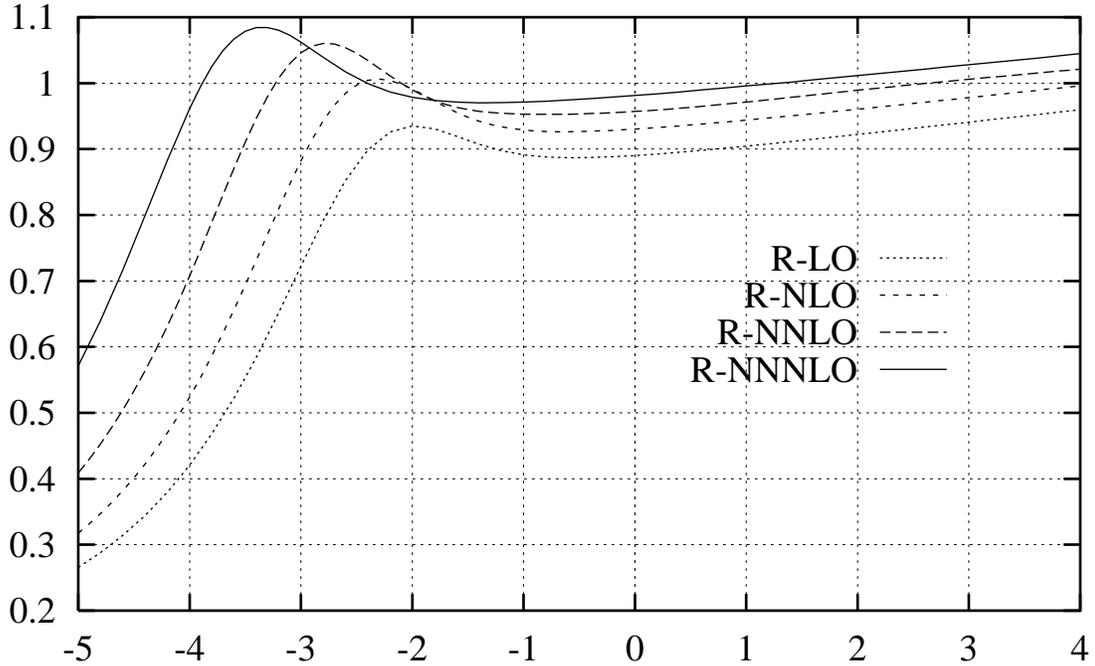}
}
\caption{\it
$R(e^+e^-\to t\bar t)$ for the LO, NLO, NNLO, NNNLO potentials
in the scheme with the running coupling
as a function of E, GeV.
We use $m_t=$175 GeV, $\alpha_s(m_Z)=$0.118, $\Gamma_t=$1.43 GeV.}
\end{figure}
In Fig.2, we plot the cross section $R$ as a function of the energy
$E$.
We have chosen  $m_t=175$ GeV, $\alpha_s(m_Z)=$0.118, $\Gamma_t=1.43$ GeV.
For a conservative estimate we have chosen
$a_3=100a_2$ (recall that $a_2 \approx 70a_1$).

Unfortunately, the resummation of the logarithms does not place the higher
order corrections under control. One can see from Fig.2, that the
corrections to the cross section are not small in the vicinity of the peak.
Moreover, these corrections do not show a decrease in
their values with increasing order. The weak convergence of the
perturbative series is caused by the enormously large non--logarithmic
coefficients, $a_n$, entering the potential (\ref{V}).
We see that the $1S$ peak is shifted by about 0.5 GeV upon
inclusion of each new order into the potential.
At least partially, this effect can be due to the use of the pole mass.
It is known that the latter suffers from the renormalon ambiguities
and thus cannot be determined with accuracy better than $\Lambda_{\rm QCD}$ 
\cite{Bigi1,Bigi2,Braun}.
The same order ambiguities are present in the interaction potential,
$V(r)$. As was shown in \cite{Ural,KU,Hoang, Beneke},
the leading order ${\cal O}(\Lambda_{\rm QCD}r)$ ambiguities cancel
in the combination $V(r) +2m$.

It is rather natural to anticipate that a more safe way to construct an
expansion
is to use a
running mass at the low normalization point $\mu$, but
the issue requires a more detailed study.

Let us now comment on the accuracy of results for the cross section.
Fig.2 shows, that the conservative estimate of $a_3$ suggests
the deviation of the NNLO curve by about $10\%$ at the peak.
We have also calculated the cross section with the NNLO
potential in the scheme with the fixed normalization point and have
compared it with the ``running'' NNLO result from Fig.2.
The difference is again about $10\%$.
Our estimate of the accuracy confirms the
conclusion of the paper \cite{Kuhn}.

{\bf 5.
}
We have studied QCD radiative corrections,
originating from the running of the coupling $\alpha_s$,
to the total cross section of the process
$e^+e^-\to t \bar t$ near the threshold.
We have resummed potentially  large
${\cal O}\Big( (\beta_0\alpha_s)^n\Big)$
short-distance QCD corrections to the production
vertex $\gamma^* t \bar t$.
The exact in $\beta_0\alpha_s$ short--distance correction
renormalizes the current to $-9.5\%$
with respect
to the Born result and is well under control.
The long distance corrections are accounted for
by using a potential with the running QCD coupling
and are more significant.
We demonstrate that the accuracy of present predictions
for the cross section in the scheme with the running QCD coupling
is about 10\%.
\newpage
{\bf Acknowledgments}

We are grateful to A.Hoang, V.Fadin, K.Melnikov and N.Uraltsev for useful 
advices. 
The work of O.Ya. was supported by the German Federal Ministry for
Research and Technology (BMBF) under contract number 05 7WZ91P (0).
A.Y. acknowledges the financial support from the Russian
Foundation for Fundamental Research, grant 98-02-17913.

\end{document}